\documentclass[conference]{IEEEtran}
\IEEEoverridecommandlockouts

\usepackage{cite}
\usepackage{amsmath,amssymb,amsfonts}
\usepackage{algorithm,algpseudocode}
\usepackage{makecell}
\usepackage{tabularx}
\usepackage{booktabs}
\usepackage{balance}
\usepackage[group-separator={,},group-minimum-digits=4]{siunitx}
\usepackage{graphicx}
\usepackage{subcaption}
\usepackage{textcomp}
\usepackage{xcolor}
\usepackage{hyperref}
\usepackage{tikz}
\usepackage{pgfplots, pgfplotstable}
\usepackage{float}
\usepackage{multirow}
\usepackage{placeins}
\newcommand{\mA}{\mathbf{A}} 
\newcommand{\mB}{\mathbf{B}}

\newcommand{\mC}{\mathbf{C}}

\newcommand{\InsetPlot}[5][0.24\textwidth]{%
\begin{subfigure}{#1}
  \centering
  \begin{tikzpicture}
    \node[anchor=south west, inner sep=0] (main) at (0,0)
      {\includegraphics[width=\linewidth]{#2}};

    \node[
      anchor=north west,
      xshift=15pt,
      yshift=-5pt,
      draw,
      line width=0.3pt,
      fill=white,
      inner sep=0      
    ] at (main.north west)
      {\includegraphics[width=0.25\linewidth]{#3}};
  \end{tikzpicture}
  \caption{#4}
  \label{#5}
\end{subfigure}%
}

\graphicspath{{figures/}}
\usetikzlibrary{calc}

\pgfplotsset{compat=1.9}

\def\BibTeX{{\rm B\kern-.05em{\sci\kern-.025em b}\kern-.08emT\kern-.1667em\lower.7ex\hbox{E}\kern-.125emX}}

\begin{document}

\title{Sparsity-Aware Roofline Models for \\ Sparse Matrix-Matrix Multiplication
}

\author{\IEEEauthorblockN{
    Matthew Qian\IEEEauthorrefmark{1},
    Yahia Ramadan\IEEEauthorrefmark{2},
    Suhita  Anubha\IEEEauthorrefmark{3},
    Ariful Azad\IEEEauthorrefmark{4} }  
  \IEEEauthorblockA{\IEEEauthorrefmark{1}
    Texas A\&M University, TX, USA 
    (mqian01@tamu.edu)}
     \IEEEauthorblockA{\IEEEauthorrefmark{2}
    Texas A\&M University, TX, USA (yahiaramadan@tamu.edu)}
    \IEEEauthorblockA{\IEEEauthorrefmark{3}
    Indiana University, Bloomington, IN, USA (sanubha@iu.edu)}
     \IEEEauthorblockA{\IEEEauthorrefmark{4}
    Texas A\&M University, TX, USA (ariful@tamu.edu)}
}

\maketitle

\begin{abstract}
Sparse matrix–dense matrix multiplication (SpMM) is a critical kernel in scientific computing, graph analytics, and machine learning, whose performance is often constrained by memory bandwidth. In this work, we investigate the applicability and limitations of roofline modeling for SpMM by explicitly accounting for the impact of matrix sparsity structure on arithmetic intensity and attainable performance. We evaluate three SpMM implementations: Compressed Sparse Row (CSR), Compressed Sparse Blocks (CSB), and Intel's Math Kernel Library (MKL). Each implementation was tested using large-scale matrices from the SuiteSparse collection and grouped by sparsity pattern, including block-structured, banded (diagonal), scale-free, and uniformly random matrices. We derive sparsity-aware roofline models that incorporate memory traffic, cache locality, and blocking behavior, and demonstrate that a single model is insufficient to accurately predict performance across diverse structures. Experiments were conducted on an AMD-based Perlmutter compute node with a varying number of columns in the dense matrix. In particular, blocking and structured sparsity significantly alter effective arithmetic intensity. The results show that accurate roofline-based performance analysis of SpMM requires sparsity-aware modeling, and that data layout and blocking strategies must be evaluated in the context of matrix structure rather than through a single unified model.
\end{abstract}

\section{Introduction}
Sparse matrix–dense matrix multiplication (SpMM) is a fundamental operation in sparse linear algebra~\cite{kepner2016mathematical, azad2021combinatorial}. It plays a central role in Graph Neural Networks (GNNs), supporting both forward and backward propagation during training and inference \cite{huang2020ge}. Beyond GNNs, SpMM underpins a wide range of graph analytics, including graph embedding \cite{rahman2020force2vec}, graph visualization \cite{rahman2020batchlayout}, and batched PageRank computations \cite{hou2021massively}. In scientific computing, SpMM is widely used in modal analysis for the Finite Element Method (FEM) \cite{gutknecht2007block} and electronic structure simulations based on Density Functional Theory (DFT) \cite{goedecker1999linear}. More recently, SpMM has emerged as a key primitive in traditionally dense domains, such as sparsified Transformers for large language models (LLMs) \cite{wang2025generalsparse} and sparsified convolutional filters in deep neural networks \cite{gale2020sparse}.

Given the broad and growing reliance on SpMM, accurate performance models are needed to predict its attainable performance on modern multicore architectures. The Roofline model~\cite{williams2009roofline} is one of the most widely used frameworks for performance analysis that relates achievable performance to arithmetic intensity, defined as the ratio of floating-point operations to memory traffic. Operations with low arithmetic intensity are classified as memory-bound, with performance primarily limited by memory bandwidth. 
While the Roofline model has proven effective for dense computations and simpler sparse kernels such as sparse matrix–vector multiplication (SpMV), its application to SpMM has received comparatively little attention.

SpMM is widely regarded as a memory-bound operation whose performance is strongly influenced by the sparsity structure of the matrix. For example, matrices with block or clustered sparsity often achieve significantly higher performance than matrices with uniformly random nonzeros. Consequently, a meaningful Roofline analysis of SpMM must explicitly account for sparsity structure. However, most existing Roofline-based models for SpMM either ignore sparsity or are tailored to specific application contexts, such as nuclear reactor simulations~\cite{aktulga2014optimizing} or sparse attention mechanisms in LLMs~\cite{shinn2023sparsity}, and thus fail to capture the wide performance variability observed across different sparsity patterns.

In this paper, we argue that no single Roofline model can adequately characterize SpMM performance across diverse sparse matrices. Instead, we propose a set of sparsity-aware Roofline models that reflect distinct structural regimes. We develop four models by deriving arithmetic-intensity bounds for random sparsity, block-diagonal sparsity, general block-structured sparsity, and scale-free sparsity. We show that random sparsity represents a worst-case scenario, providing a lower bound on SpMM performance, while structured sparsity enables significantly higher attainable performance. 

We validate our models using four classes of matrices corresponding to each sparsity structure and evaluate three SpMM implementations based on compressed sparse row (CSR), compressed sparse block (CSB)~\cite{csb}, and the Intel MKL library. Experimental results on an AMD multicore processor demonstrate that sparsity-aware Roofline models closely track observed performance. These results confirm that incorporating sparsity structure is essential for accurately predicting SpMM performance.

Overall, the paper makes the following contributions. 
\begin{enumerate}
    \item We introduce sparsity-aware Roofline models for SpMM that explicitly capture the impact of sparsity structure. We derive the arithmetic intensity and modeling for random, diagonal, scale-free, and block-structured matrices. 

    \item We validate the proposed models by showing that sparsity-aware Roofline predictions closely match measured performance across CSR, CSB, and MKL SpMM implementations.
\end{enumerate}

\begin{table}[!t]
\centering
\caption{Notations used in the paper.}
\begin{tabular}{ll}
\toprule
\textbf{Symbol} & \textbf{Description} \\
\toprule
$\mA$ & Sparse matrix of size $n \times n$ \\
$\mB$ & Dense matrix of size $n \times d$ (tall-and-skinny) \\
$\mC$ & Output matrix of size $n \times d$ (tall-and-skinny) \\
$n$ & Number of rows/columns in $\mA$ \\
$d$ & Number of columns in $\mB$ (and $\mC$) \\
$\text{nnz}$ & Number of nonzeros in $\mA$ \\
$t$ & Block dimension for $\mA$ \\
$D$ & Average number of nonzeros in a block of $\mA$  \\
$z$ & Average number of nonempty columns in a block of $\mA$ \\
\midrule
$\text{FLOPs}$ & Total floating point operations in SpMM ($\approx 2 \cdot \mathrm{nnz} \cdot d$) \\
AI & Arithmetic intensity ($\text{AI} = \text{FLOPs} / \text{Bytes}$) \\
$\beta$ & Peak memory bandwidth (122.6 GB/s) \\
\bottomrule
\end{tabular}
\label{tab:notation}
\end{table}

\begin{table*}[!t]
    \centering
    \caption{ Common matrix shapes and sparsity patterns arising in SpMM applications across machine learning and scientific computing.}
    \begin{tabular}{llllll}
    \toprule
     & & $\mA$ & & $\mB$  \\
     Applications & Type & Shape & Common Sparsity &  Type & Shape  \\
    \toprule
    GNN and graph embedding &  adjacency matrix &  square & scale-free & node features & tall-and-skinny \\ 
    Finite Element Method (FEM) &  stiffness/mass matrices &  square & banded, mesh-local & eigenvectors & tall-and-skinny\\ 
    Density Functional Theory (DFT) & Hamiltonian & square & banded & Kohn–Sham orbitals & tall-and-skinny\\ 
    Transformer Sparsification & sparse attention map & square & top-k/row, banded & dense value matrix & tall-and-skinny\\ 
    Recommendation Systems & user-item matrix & rectangular & irregular & item embeddings & tall-and-skinny\\ 
    \bottomrule
    \end{tabular}
    \label{tab:SpMM_App}
\end{table*}

\section{Background and Related Work}

SpMM multiplies a sparse matrix 
$\mA \in \mathbb{R}^{n \times n}$ with a dense matrix 
$\mB \in \mathbb{R}^{n \times d}$ to produce a dense matrix 
$\mC \in \mathbb{R}^{n \times d}$. 
Table~\ref{tab:notation} outlines the notation used throughout this paper.
The performance of SpMM depends on multiple factors, including matrix dimensions, sparsity structure, data layout, implementation strategy, and the underlying hardware architecture. 
In this section, we discuss these factors to provide the necessary context and motivate the design of our Roofline performance models.

\subsection{Matrix Characteristics}
The sizes of matrices affect cache usage and thereby impact the model and performance. 
Table~\ref{tab:SpMM_App} shows five applications of SpMM in scientific computing and machine learning. 
All of these applications use a square sparse matrix and a tall-and-skinny dense matrix where $d{\ll} n$.

In addition, the distribution of nonzeros strongly influences SpMM performance. Random sparsity leads to irregular memory accesses and poor cache reuse, while banded structures improve locality by reusing portions of $\mB$ across multiple multiplications. 
Higher average density increases arithmetic intensity and increases theoretical performance, whereas very sparse rows or columns increase bandwidth pressure.

\begin{center}
\fbox{
\parbox{0.95\linewidth}{
\textbf{Modeling Assumptions.} 
Based on the above discussion, we restrict our analysis to \emph{tall-and-skinny} dense matrices. 
To account for the impact of sparsity structure on performance, we develop models for random, banded, block, and scale-free sparsity patterns.
}
}
\end{center}

\subsection{Algorithm and Data Layout}
Matrix format significantly affects SpMM performance. Formats like CSR, CSC, and CSB differ in how they store and access nonzeros, impacting cache locality and memory traffic. Blocking or tiling strategies, as in CSB, enhance cache reuse by working on submatrices, improving performance on modern memory hierarchies. Additionally, the choice of multiplication algorithms such as column-by-column, inner product, or outer product affects data access patterns and parallelism, influencing efficiency and scalability.

\begin{center}
\fbox{
\parbox{0.95\linewidth}{
\textbf{Data Layout Options.} 
We consider SpMM implementations based on the CSR, CSC, and CSB data structures to capture commonly used sparse formats and enable fair and representative performance modeling and comparisons.
}
}
\end{center}

\subsection{The Roofline model}
The Roofline model~\cite{williams2009roofline} is a visually intuitive performance framework that relates a kernel’s attainable performance to the architectural limits of a computing platform. 
By plotting performance (GFLOP/s) against arithmetic intensity (AI), the model identifies whether a kernel is ``memory-bound," limited by the hardware's peak bandwidth, or ``compute-bound," limited by its peak computational throughput. More precisely, arithmetic intensity is defined as the ratio of the total number of floating-point operations performed by a kernel to the total number of bytes of data transferred to and from the main memory during its execution ($AI = \frac{\text{FLOPs}}{\text{Bytes}}$). This metric quantifies the ``computational density" of an algorithm; a higher AI suggests greater data reuse within local caches, allowing a kernel to potentially reach the hardware's maximum processing speed rather than being throttled by the slower memory subsystem.

The two bounding sections are defined by $P = \min(\beta\cdot AI, \pi)$,
where $\pi$ is the peak compute throughput, and $\beta$ is the peak memory bandwidth.
For SpMM, the kernel is predominantly memory-bound. This is because each nonzero in the sparse matrix contributes relatively few floating point operations (on the order of $2d$ FLOPs) while requiring multiple memory accesses. As a result, the arithmetic intensity remains low, placing SpMM in the memory-bound region of the roofline model.
In this regime, performance scales linearly with arithmetic intensity and is therefore bounded by the sloped line $P = \beta \cdot AI$. Only when the arithmetic intensity exceeds the ridge point ($AI > \pi / \beta$) does the kernel become compute-bound, at which point performance is capped by $\pi$.

\subsection{Related Work}
To better reflect modern architectures, several extensions incorporate hierarchical bandwidth limits across cache levels, yielding cache-aware or hierarchical Roofline models that improve diagnostic power for locality-sensitive kernels~\cite{ilic2013cache}.

Developing a Roofline model that captures the observed performance of sparse operations is more challenging than for dense kernels. This is because the standard Roofline model does not account for locality, sparsity structure, or indirect memory accesses.
Prior work has used Roofline-style reasoning to analyze sparse linear algebra kernels such as SpMV~\cite{williams2007optimization}, SpGEMM~\cite{gu2020bandwidth}, and FusedMM~\cite{rahman2021fusedmm}, often concluding that irregular access patterns and limited locality reduce effective bandwidth well below hardware peaks~\cite{aktulga2014optimizing}. 

SpMM occupies an intermediate position between SpMV and dense GEMM: while it inherits indirect accesses from the sparse operand, it can exploit reuse through the dense matrix, making its performance highly sensitive to matrix structure and dense-block shape. Several application-driven studies analyze SpMM and related kernels using Roofline-style models, both in scientific computing~\cite{aktulga2014optimizing} and emerging machine-learning workloads such as sparse attention~\cite{shinn2023sparsity}. 

Optimized SpMM is a widely studied problem, with algorithms developed for multicore processors~\cite{hoque2024isplib}, GPUs~\cite{huang2020ge, hong19adaptivetiling, li22quantizedspmm}, and distributed systems~\cite{ics21spmm, block2024two, koanantakool2016communication}. 
Works such as  WISE~\cite{yesil23_wise} and DDB~\cite{yesil22ddb} attempt to optimize sparse computations by using machine learning techniques to predict the performance of various configurations.
However, the performance of these algorithms is rarely modeled analytically, making it difficult to assess whether the observed performance approaches the peak achievable performance of the target platform.







\section{Sparsity-Aware Roofline models for SpMM}
\label{sec:roofline}
To develop sparsity-aware roofline models, we consider $n\times n$ square sparse matrix and $n\times d$ tall-and-skinny dense matrix.
Since we only perform a multiply-accumulate for each non-zero element in the sparse matrix $\mA$ for every column ($d$) in the dense matrix $B$:
\begin{equation}
    \text{FLOP} = 2d \cdot \text{nnz}.
\end{equation}
The factor of 2 accounts for one multiplication and one addition per non-zero.

To estimate memory traffic, we assume that all matrix values are stored in double-precision floating-point format, while indices in the sparse matrix are stored as 32-bit integers. The total memory traffic is computed as the sum of the costs of reading $\mA$, reading $\mB$, and writing $\mC$.

 For $\mA$  in CSR format, we must read the values, column indices, and row pointers. Thus, total data movement for $\mA$ is: 
 \[
\begin{aligned}
\text{Traffic}_A &= \text{nnz} \cdot \text{sizeof(val)} + \text{nnz} \cdot \text{sizeof(col\_idx)} \\
&\ \ \ \ \ + (n+1) \cdot \text{sizeof(row\_ptr)} \\
&\approx 12\text{nnz bytes} 
\end{aligned}
\]
 
Since $\mC$ is dense, it is typically written once, giving $\text{Traffic}_C = nd \cdot \text{sizeof(val)} = 8nd$ bytes.
The remaining question, therefore, is how $\mB$ is accessed under different sparsity patterns of the sparse matrix. In the following, we consider four representative sparsity patterns and develop a corresponding access model for $\mB$ in each case.

\subsection{Random Sparsity}
Under a random sparsity pattern, the column indices of $\mA$ are assumed to be uniformly distributed. If the dense matrix $\mB$ does not fit in cache, each nonzero of $\mA$ incurs a memory access to load the corresponding row of $\mB$ from main memory, resulting in no effective data reuse for $\mB$.
Hence, in the worst case, we access nnz rows of $\mB$ for the entire multiplication.

Thus, total data movement for $\mB$ in the random sparsity model is: 

 \[
\begin{aligned}
\text{Traffic}_B &=  \text{nnz} \cdot d \cdot \text{sizeof(val)}  \\
&= 8d \cdot \text{nnz bytes} 
\end{aligned}
\]


Using the previously defined FLOP count, and $\text{Traffic}_A$ and $\text{Traffic}_C$, we can derive the arithmetic intensity for the random sparsity: 
\begin{align}
\text{AI (Random)} &= \frac{2d \cdot \text{nnz}}{(12 + 8d)\cdot \text{nnz} + 8nd}
\end{align}








\subsection{Diagonal Sparsity}

For Diagonal sparsity, we assume that the nonzeros of $\mA$ are concentrated along the main diagonal. More concretely, we assume that the number of nonzeros per row is small and bounded, and that the nonzeros have strong spatial locality between consecutive rows. 
This assumption allows access to $\mB$ to have high temporal locality, such that rows that are accessed once in $\mB$ are accessed repeatedly before being evicted from cache. Thus, with an optimal data structure, the memory traffic for $\mB$ is amortized across all nonzeros in the diagonal region and does not scale with the number of nonzeros in $\mA$. Consequently, the dominant memory traffic for this diagonal case is dependent on reading from $\mA$, writing to $\mC$, and a single load of $\mB$ into cache, so we have:

\begin{align}
\text{AI (Diagonal)} &= \frac{2d \cdot \text{nnz}}{12 \cdot \text{nnz} + 16nd}
\end{align}






\subsection{Blocked Sparsity}
Assume  $\mA \in \mathbb{R}^{n \times n}$  is divided into $t \times t$ blocks, with a total of $\text{nnz}$ nonzero elements and $N$ total nonzero blocks. Let $D = \text{nnz}/N$ be the average number of nonzero entries per nonzero block.

We assume that it takes 8 bytes to read the values, and 4 bytes to read an index for values within a block. Since we access each value in $\mA$ once, the total memory traffic for $\mA$ is simply 12 nnz. We also assume that we write 8 bytes to the dense matrix $\mC$, so the memory traffic for $\mC$ is $8nd$.

To model the memory access for $\mB$, we assume that each $t \times t$ block of $\mA$ must access $\mB$ once for each of its $t$ columns that contain a nonzero. To estimate the average number of such columns (denoted $z$) per block, we can assume that nonzeros within a single block are distributed randomly among its $t$ columns.

We define an indicator random variable $I_i$ for each column $i$ within a block.

$$I_i = \begin{cases}
    1, & \text{if col } i ~\text{{contains at least one nonzero}} \\
    0, & \text{if col } i \text{ is empty}
\end{cases}$$

Thus, the expected number of nonzeros in a block is given by
$$\mathbb{E}[z] = \sum_{i=0}^{t} \mathbb{E}[I_i] = t \cdot \text{Pr(col $i$ is nonempty)}$$


A column $i$ is empty if all $D$ nonzeros are outside of $i$. For a single nonzero, Pr(miss col $i$) = $1 - \frac{1}{t}$. So for $D$ independent nonzeros, we have

$$\text{Pr(col $i$ is nonempty)} = 1 - \text{Pr(col $i$ is empty)} = 1-(1 - \frac{1}{t})^D,$$

When $t$ (block size) and $D$ (number of nonzeros in a block) are large, we can use the following approximation:

$$(1 - \frac{1}{t})^D \approx e^{-D/t},$$

because for large $t$ and $D$, the binomial distribution converges to a Poisson distribution.

Finally, we have $z \approx \mathbb{E}[z] = t(1-e^{-D/t})$ occupied columns per block~\cite{mitzenmacher2017probability}. A row of $\mB$ is accessed for each block $z$ times, and thus the memory traffic for $\mB$ is given by the expression $8dNz$. 
Due to the cache tiling that the blocked structure provides, in practice, repeatedly accessing nearby rows or columns of $B$ allows for effective cache reuse rather than main memory access. To roughly account for this reuse, we scale the memory traffic from $B$ by a factor of $\frac{1}{4}$. This heuristic reflects that due to tiling, only part of a single theoretical memory access actually results in a main memory access. The exact fraction depends on hardware, cache behavior, and implementation, so we choose $\frac{1}{4}$ as an estimate based on observed experimental results. Including this factor in our full equation gives us: 


\begin{align}
\text{AI (Blocked)} &= \frac{2d \cdot \text{nnz}}{8\,\text{nnz} + 2dNz + 8nd}
\end{align}

\subsection{Scale-free Sparsity}
For scale-free graphs, we assume that the degree distribution $p(k)$ of a graph follows a power law distributation $p(k) \propto k^{-\alpha}$, where $\alpha$ is 
the power-law exponent with $2 < \alpha < 3$ for real-world networks~\cite{clauset2009power}.
For scale-free graphs, a small number of high-degree hub nodes can account for a disproportionately large fraction of nonzeros. Let $n_{\mathrm{hub}}$ be the number of hub nodes and $\text{nnz}_{\mathrm{hub}}$ be nonzeros (edges) associated with hub nodes. 
In the appendix, we derived the following estimation for $\text{nnz}_{\mathrm{hub}}$:
\begin{align}
\text{nnz}_{\mathrm{hub}}
&=
\text{nnz}\cdot f^{(\alpha-2)/(\alpha-1)},
\end{align}
where $f$ is the highest fraction of nodes by degree.
In our experiments, we set the hub fraction to $f$=0.1\% of the nodes.

In our model, we assume that the rows of 
$\mB$ associated with hub nodes can be retained in fast memory (e.g., cache) and therefore are not repeatedly loaded, unlike in the random sparsity model. For the remaining non-hub entries of the sparse matrix, accesses to $\mB$ are assumed to be random. Under these assumptions, the arithmetic intensity for scale-free sparsity is defined as:

\begin{align}
\text{AI} &= \frac{2d \cdot \text{nnz}}{12 \text{nnz} + 8d\cdot (\text{nnz}-\text{nnz}_{\mathrm{hub}}) + 8d\cdot n_{\mathrm{hub}}  + 8nd} 
\end{align}

\begin{table}[!t]
\centering
\caption{Sparse matrices used for SpMM performance evaluation. er\_22\_10 It is generated using the Erdős–Rényi model with $2^{22}$ rows and columns and an average of 10 nonzeros per row.}
\label{tab:matrices}
\begin{tabular}{llrrr}
\toprule
\textbf{Pattern} & \textbf{Matrix Name} & \textbf{Rows} & \textbf{Cols} & \textbf{Nonzeros} \\
\toprule
\multirow{4}{*}{Blocking}
& road\_usa          & 23,947,347 & 23,947,347 & 57,708,624 \\
& hugebubbles-00010  & 19,458,087 & 19,458,087 & 58,359,528 \\
& asia\_osm          & 11,950,757 & 11,950,757 & 25,423,206 \\
& 333SP              &  3,712,815 &  3,712,815 & 22,217,266 \\
\midrule
\multirow{3}{*}{Scale-free}
& com-Orkut          &  3,072,441 &  3,072,441 & 234,370,166 \\
& com-LiveJournal    &  3,997,962 &  3,997,962 & 69,362,378 \\
& uk-2002            & 18,520,486 & 18,520,486 & 298,113,762 \\
\midrule
\multirow{2}{*}{Diagonal}
& rajat31            &  4,690,002 &  4,690,002 & 20,316,253 \\
& ideal\_diagonal\_22 &  4,194,304 &  4,194,304 & 4,194,304 \\
\midrule
\multirow{3}{*}{Random}
& er\_22\_1          &  4,194,304 &  4,194,304 & 4,194,304 \\
& er\_22\_10         &  4,194,304 &  4,194,304 & 41,942,990 \\
& er\_22\_20         &  4,194,304 &  4,194,304 & 83,885,880 \\
\bottomrule
\end{tabular}
\end{table}

\begin{table}[!t]
\centering
\caption{Test system CPU specifications (Perlmutter)}
\label{tab:cpu_specs}
\begin{tabular}{ll}
\toprule
\textbf{Property} & \textbf{Value} \\
\toprule
Architecture        & x86\_64 \\
CPU model           & AMD EPYC 7763 (Milan) \\
Sockets             & 2 (testing restricted to 1 socket) \\
Cores / Threads     & 64 cores / 64 threads (used) \\
L1 cache            & 64 KiB per core (32 KiB data, 32 KiB instruction) \\
L2 cache            & 512 KiB per core \\
L3 cache            & 256 MiB per socket \\
Vector extensions   & AVX2, FMA \\
Memory              & 512 GB DDR4 total \\
NUMA domains        & 4 per socket (NPS=4) \\
\bottomrule
\end{tabular}
\end{table}

\begin{table*}[!t]
\centering
\caption{SpMM performance (GFLOP/s) for multiple $k$ values across different formats.}
\label{tab:spmm_k_multicol}
\renewcommand{\arraystretch}{1.2} 
\setlength{\tabcolsep}{3pt} 
\begin{tabular}{ll
    ccc ccc ccc ccc}
    \toprule
    \multirow{2}{*}{\textbf{Pattern}} & \multirow{2}{*}{\textbf{Matrix}} &
    \multicolumn{3}{c}{$d=1$} & \multicolumn{3}{c}{$d=4$} &
    \multicolumn{3}{c}{$d=16$} & \multicolumn{3}{c}{$d=64$} \\
    \cmidrule(lr){3-5} \cmidrule(lr){6-8} \cmidrule(lr){9-11} \cmidrule(lr){12-14}
    & & \textbf{CSR} & \textbf{MKL} & \textbf{CSB}
      & \textbf{CSR} & \textbf{MKL} & \textbf{CSB}
      & \textbf{CSR} & \textbf{MKL} & \textbf{CSB}
      & \textbf{CSR} & \textbf{MKL} & \textbf{CSB} \\
    \midrule
    \multirow{2}{*}{Blocking}
     & road\_usa  & 9.468 & 11.0924 & 14.240 & 17.528 & 17.289 & 25.423 & 32.768	& 32.652 & 36.234 & 41.316 & 38.567 & 43.006 \\
     & hugebubbles-00010 & 5.875 & 7.146	& 9.696 & 14.358 & 13.490 & 15.853 & 21.743 & 22.975 & 28.322 & 21.743 & 22.975 & 28.322  \\
     & asia\_osm & 7.301 & 10.078 & 10.668 & 20.455 & 21.481 & 14.027 & 33.975 & 34.568 & 35.093 & 38.345 & 38.450 & 33.479  \\
     & 333SP  & 5.284 & 8.692 & 13.057 & 12.258 & 23.625 & 24.875 & 28.784 & 	28.893 & 35.227 & 29.729 & 30.106 & 39.596 \\
    \midrule
    \multirow{1}{*}{Scale-free}
     & com-Orkut    & 8.402 & 18.340 & 26.894 & 14.505 & 30.560 & 38.501 & 21.037	& 29.053 & 34.403 & 12.256 & 22.460 & 32.017 \\
     & com-LiveJournal  & 11.536	& 15.010 & 26.984 & 35.687 & 44.851	& 72.008 &  66.266 & 76.981	& 92.091 & 41.683 & 53.544 & 61.322 \\
     & uk-2002         & 16.701 & 24.139	& 16.204 & 55.851 & 78.538 & 67.526 & 146.583	& 167.960 & 148.299 & 226.757 & 205.945	& 164.359 \\
    \midrule
    \multirow{1}{*}{Diagonal}
     & ideal\_diagonal\_22       & 1.988	& 1.167	& 5.886 & 23.546 & 10.558 & 	6.840 & 8.5888 & 9.039 & 14.202 & 10.902 & 11.023 & 17.294 \\
     & rajat31          &  7.266	& 9.565	& 9.390 & 26.944 & 29.348 & 22.601 & 56.978	& 59.644 & 39.275 & 74.064	& 69.266	& 53.911 \\
    \midrule
    \multirow{1}{*}{Uniform Random}
     & er\_22\_1        & 1.586 & 1.634 & 3.998 & 4.957 & 5.446 & 6.226 & 7.841	 & 8.194 & 10.216 & 8.547 & 5.320 & 11.509 \\
     & er\_22\_10       & 6.194 & 7.833 & 12.832 & 13.921 & 15.225 & 12.373 & 12.284 & 12.374 & 13.456 & 10.0322 & 11.185 & 17.036 \\
     & er\_22\_20       & 8.091 & 10.906	& 16.283 & 14.979 & 16.249 & 15.453 & 13.575 & 14.169 & 13.483 & 11.564 & 10.429	& 17.001 \\
    \bottomrule
\end{tabular}
\end{table*}

\section{Evaluating SpMM Implementations using Sparsity-Aware Roofline Models}

\subsection{Dataset}
For data collection, matrices were collected and categorized according to four structures: blocking, scale-free, diagonal, and uniform random. 
All matrices were obtained from the SuiteSparse Matrix Collection~\cite{davis2011university}, except for the uniform random matrices, which were generated using the Erdős–Rényi model, and \texttt{ideal\_diagonal\_22}, which represents an ideal diagonal sparsity pattern.
Table~\ref{tab:matrices} lists the details of each matrix used. All matrices were selected to exceed the capacity of on-chip caches. This choice allows us to isolate the main weakness of the memory-bound operation by analyzing cache utilization across different implementations.

\subsection{Experimental Settings}

To evaluate performance, we used written benchmarks that directly measure the execution time of the SpMM operation. All code was compiled using the GNU C++ compiler, with Intel oneAPI to link MKL and OpenMP for parallelization. Experiments were conducted on a single node of the Perlmutter system, and benchmarks were executed using 64 threads. Memory allocation was controlled using \texttt{numactl --interleave=all}, distributing memory pages across available NUMA domains on the node. Memory bandwidth was measured using the STREAM \cite{stream} benchmark, where we measured a peak bandwidth $\beta$ of 122.6 GB/s. Table~\ref{tab:cpu_specs} summarizes the CPU specifications of the system used for evaluation.

For measuring, only the actual SpMM operation was recorded, disregarding file loading times and variable initialization. This ensures arithmetic intensity, and the performance for the roofline model includes only floating point operations from the matrix multiplication calculation. We compared each matrix with CSR, MKL, and CSB implementations, as well as by $d$, the number of columns in $\mB$. Specifically, the values used are $d=1, 4, 16, 64$.  

The CSB implementation evaluated in this work is based on the original Cilk Plus implementation \cite{csb} and was modified and updated to support OpenMP in order to enable direct comparison for SpMM.

\definecolor{csrblue}{HTML}{4285F4}
\definecolor{mklred}{HTML}{EA4335}
\definecolor{csbyellow}{HTML}{FBBC04}

\begin{figure}[!t]
  \centering
  
  \begin{tikzpicture}
    \node[] {
      \begin{tikzpicture}[baseline]
        \draw[line width=2pt, csrblue] (0,0) -- (0.9,0);
        \node[anchor=west] at (1.05,0) {CSR};

        \draw[line width=2pt, mklred] (2.2,0) -- (3.1,0);
        \node[anchor=west] at (3.25,0) {MKL};

        \draw[line width=2pt, csbyellow] (4.4,0) -- (5.3,0);
        \node[anchor=west] at (5.45,0) {CSB};
      \end{tikzpicture}
    };
  \end{tikzpicture}

  \vspace{0.6em}

\InsetPlot{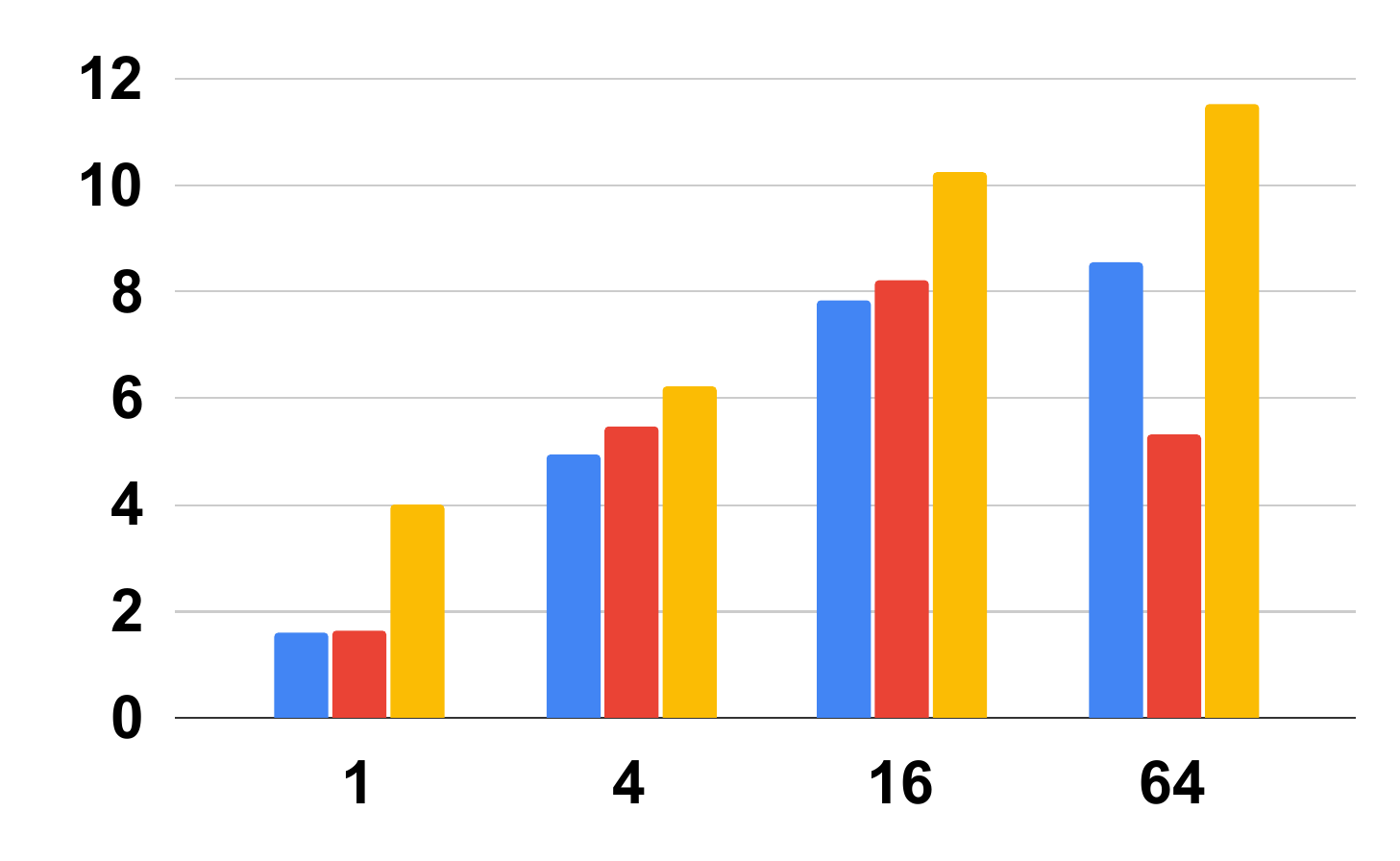}{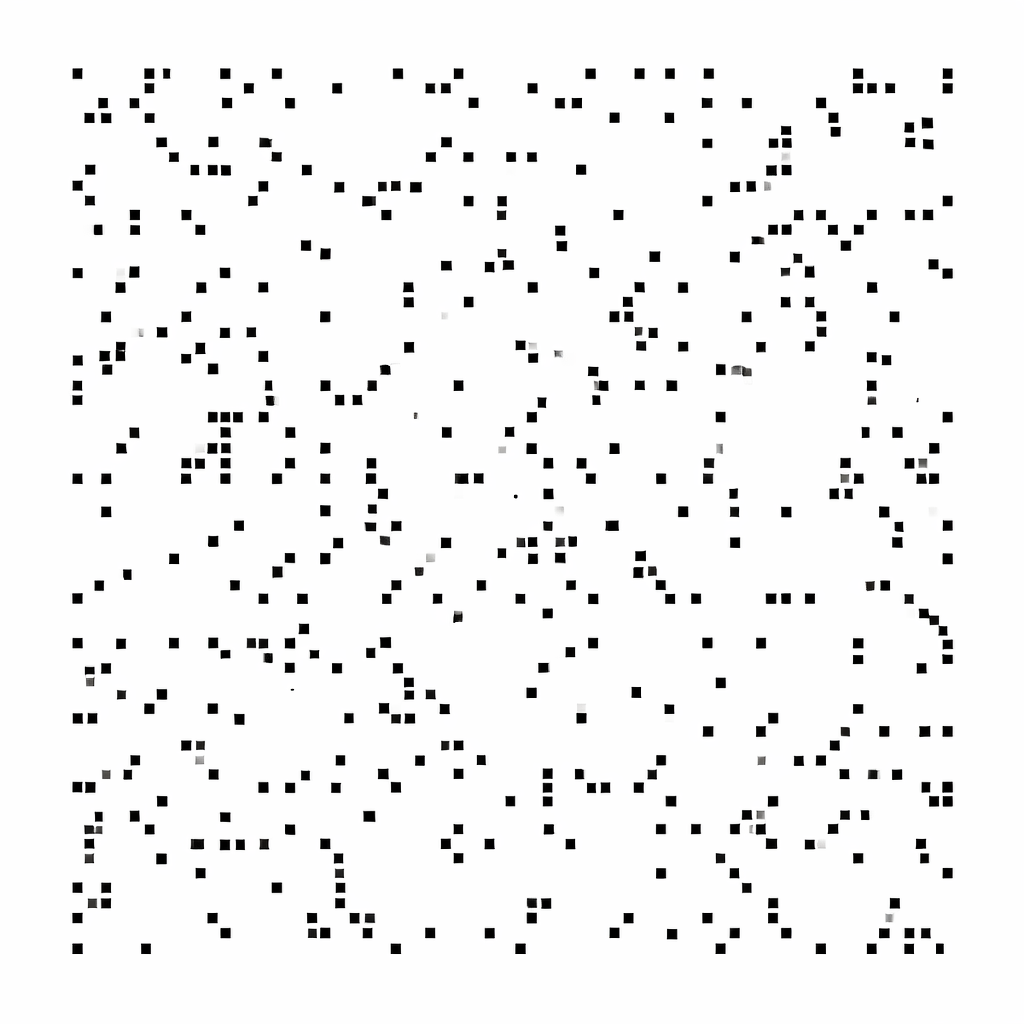}{Uniform Random}{}
\hfill
    \InsetPlot{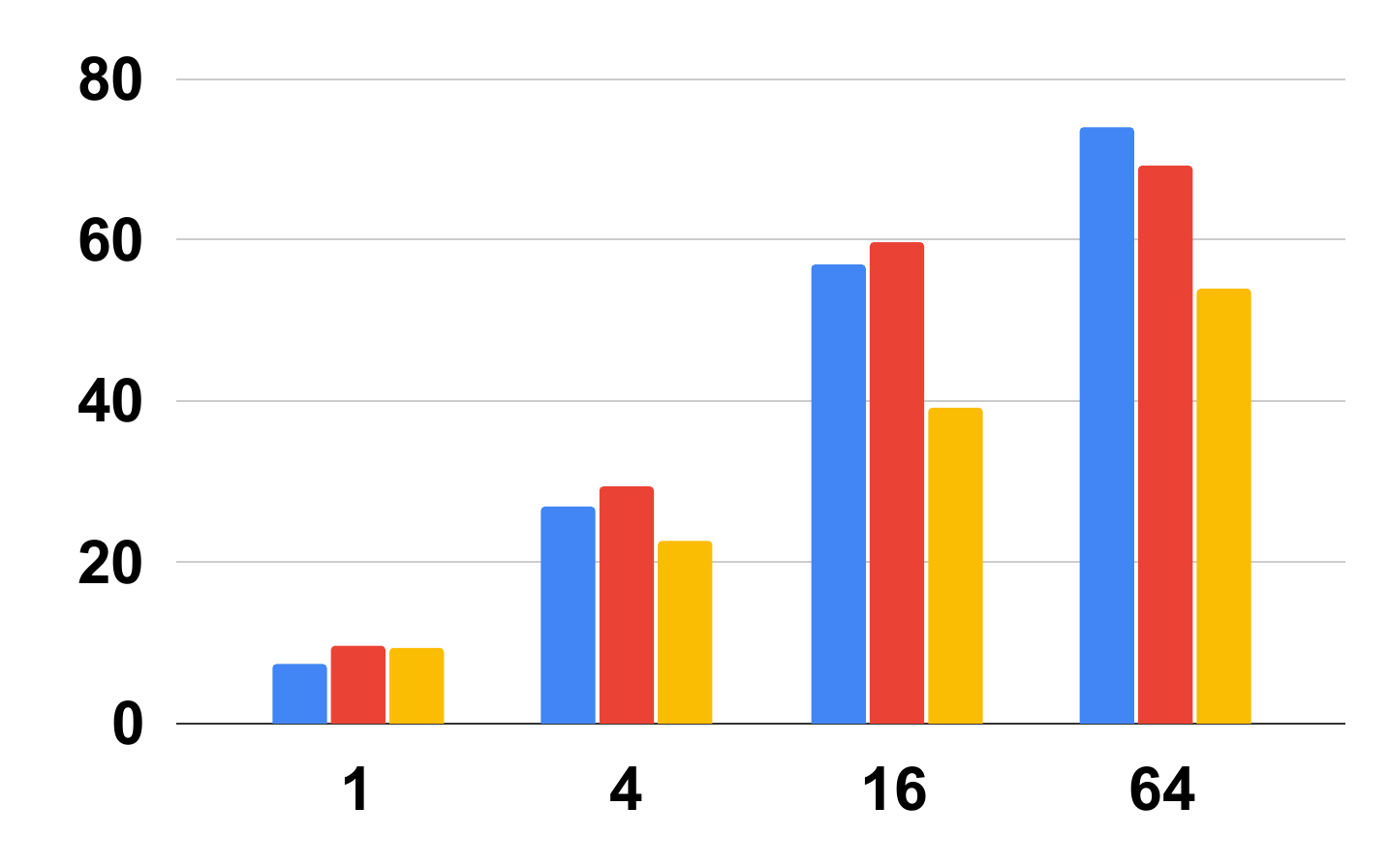}{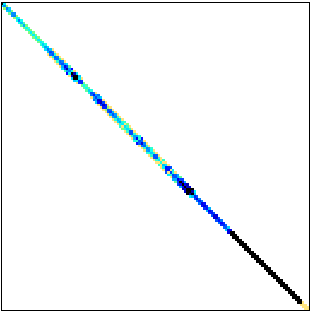}{Diagonal}{}

   \vspace{0.8em}
   
  \InsetPlot{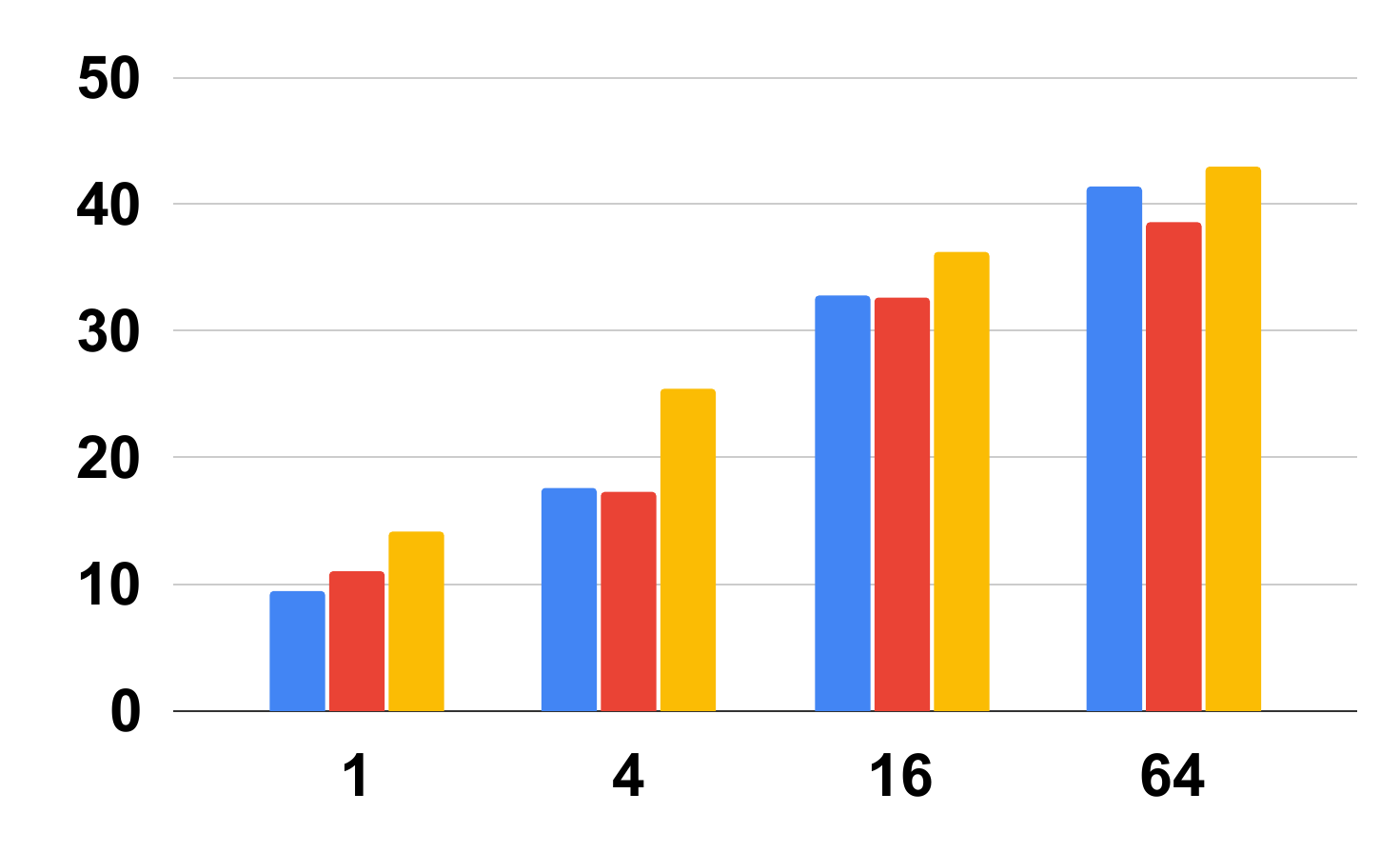}{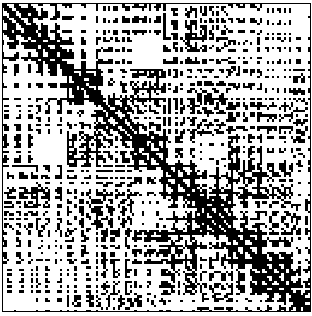}{Blocking}{}
  \hfill
  \InsetPlot{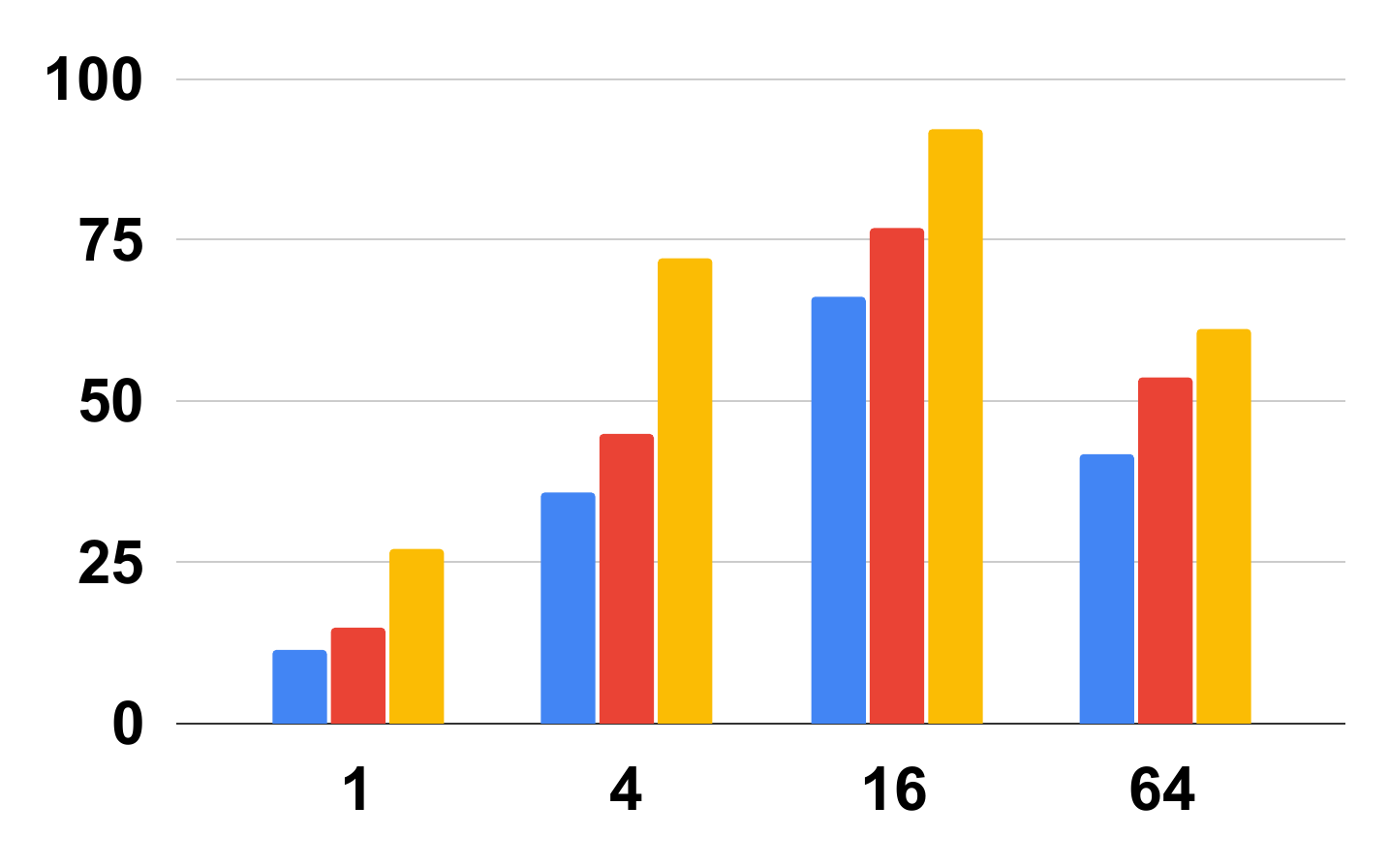}{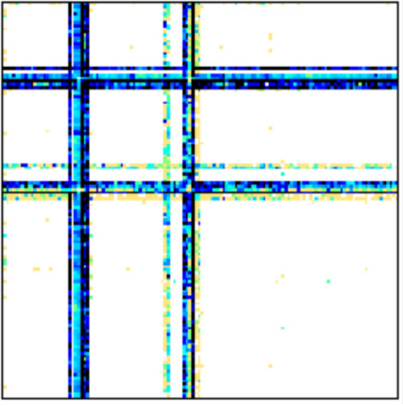}{Scale-Free}{}

  \caption{SpMM Performance (GFLOP/s) vs Number of Columns (d) for various sparsity patterns. Insets are visualizations of the specific sparsity pattern . Matrices used are (a) er\_22\_1, (b) rajat31, (c) road\_usa, and (d) com-LiveJournal.}
  \label{fig:spmm_all}
\end{figure}
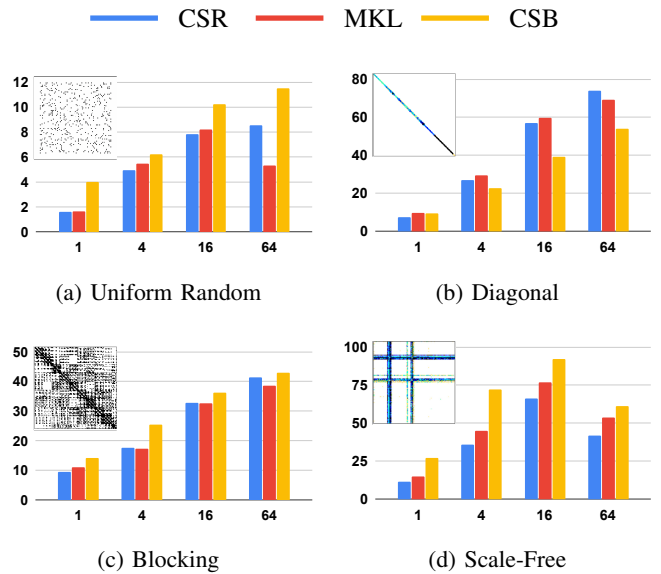

\begin{figure*}[!t]
  \centering

\includegraphics[width=0.48\textwidth]{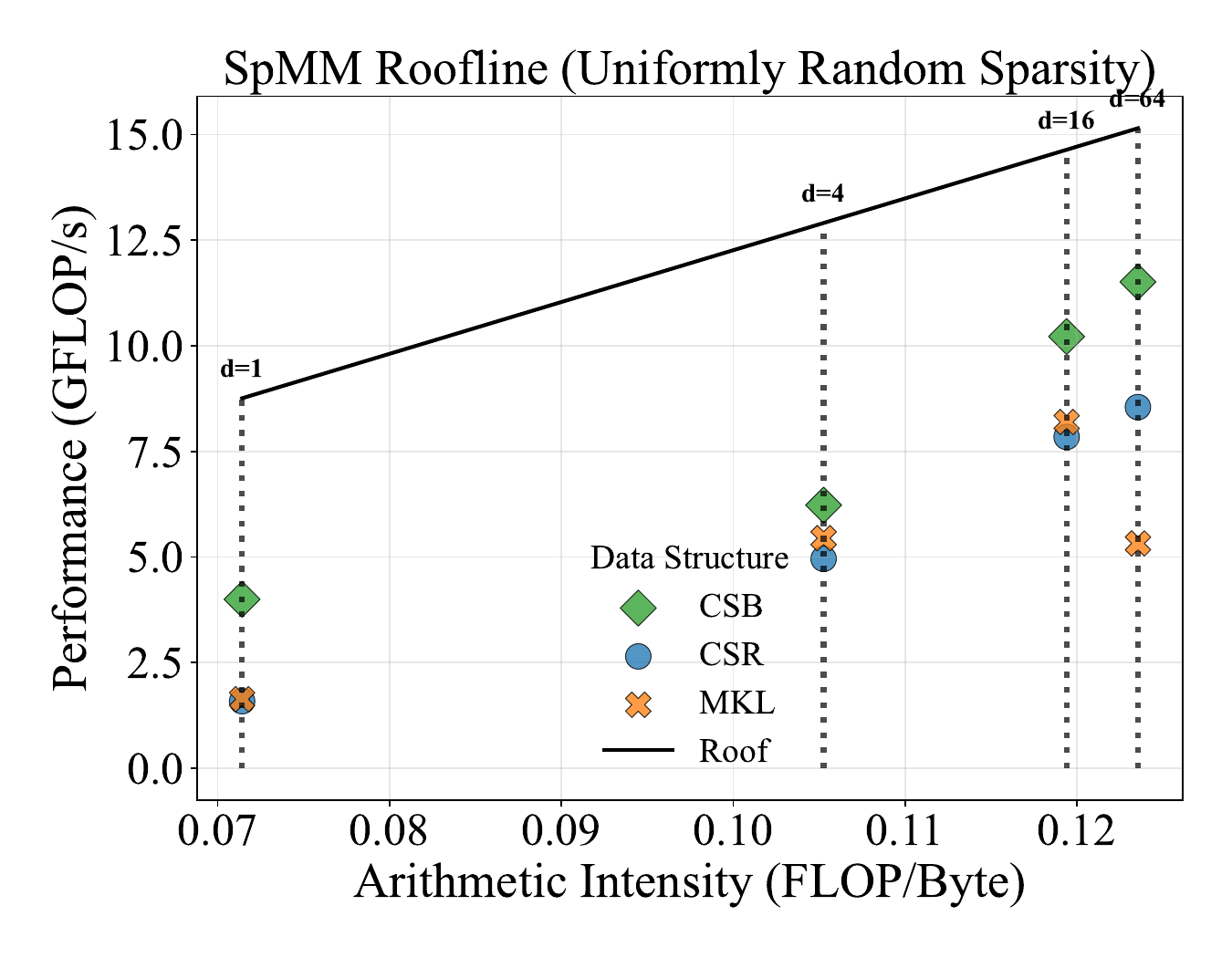}\hfill
\includegraphics[width=0.48\textwidth]{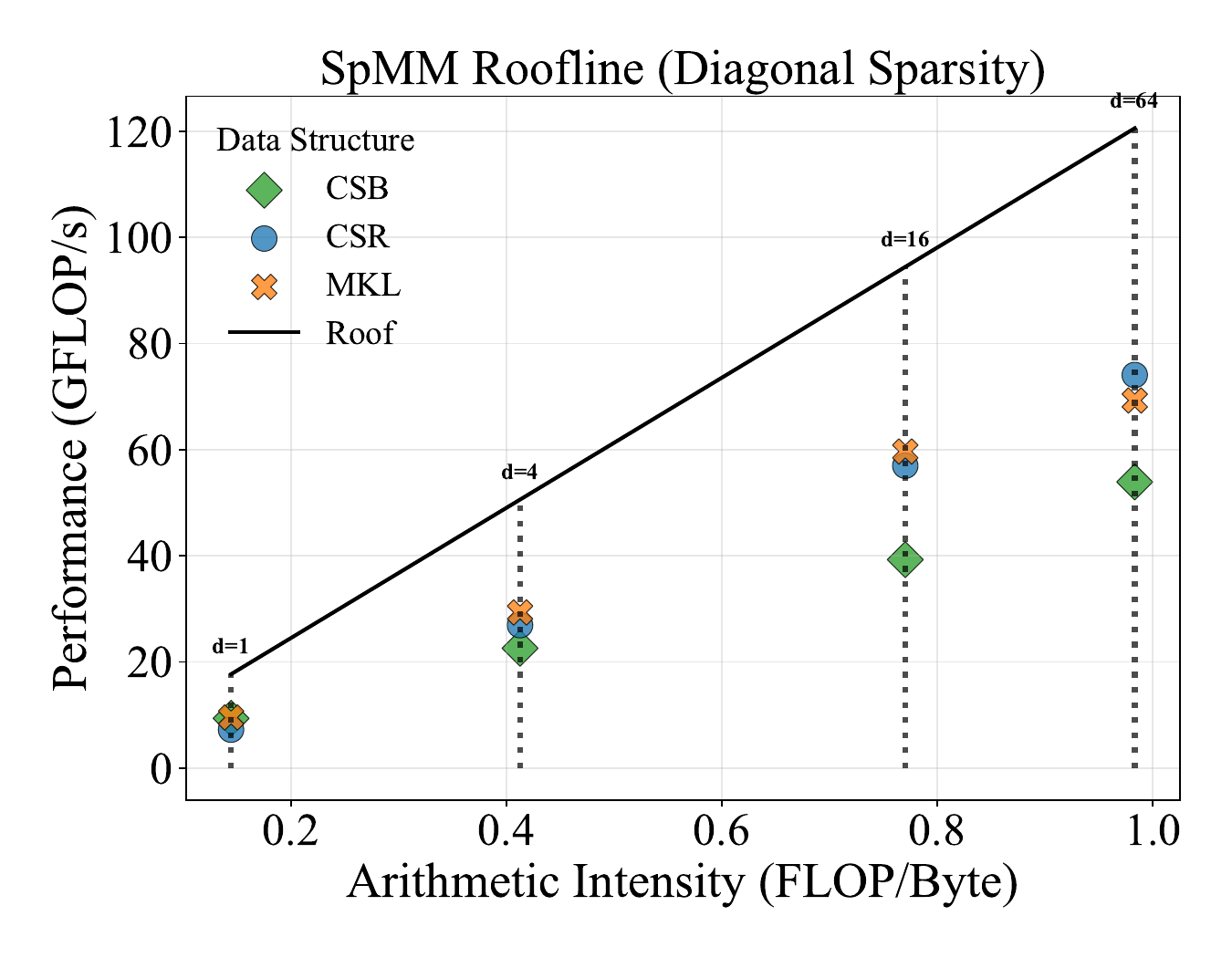}
 \vspace{0.5em}
  \includegraphics[width=0.48\textwidth]{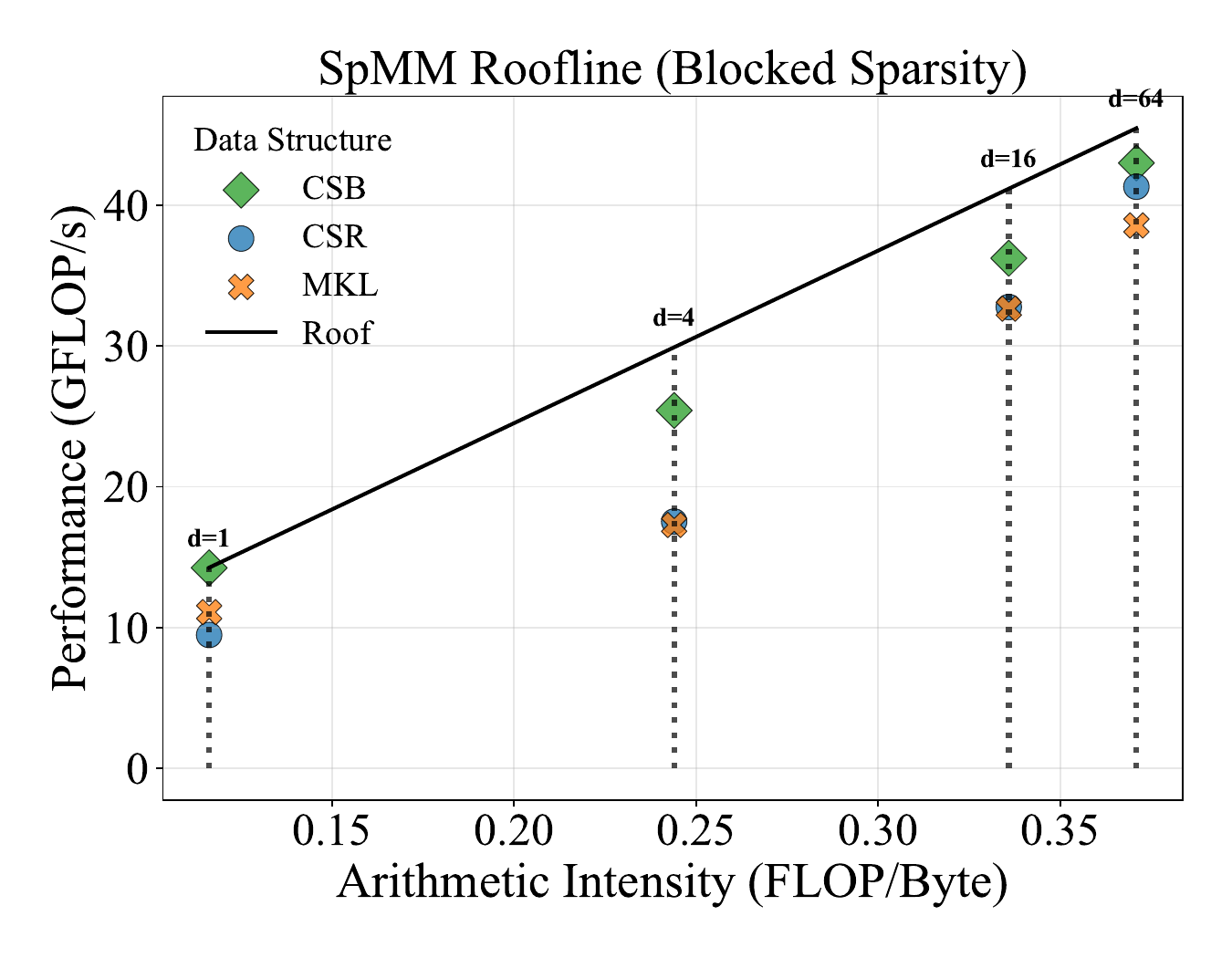}\hfill
    \includegraphics[width=0.48\textwidth]{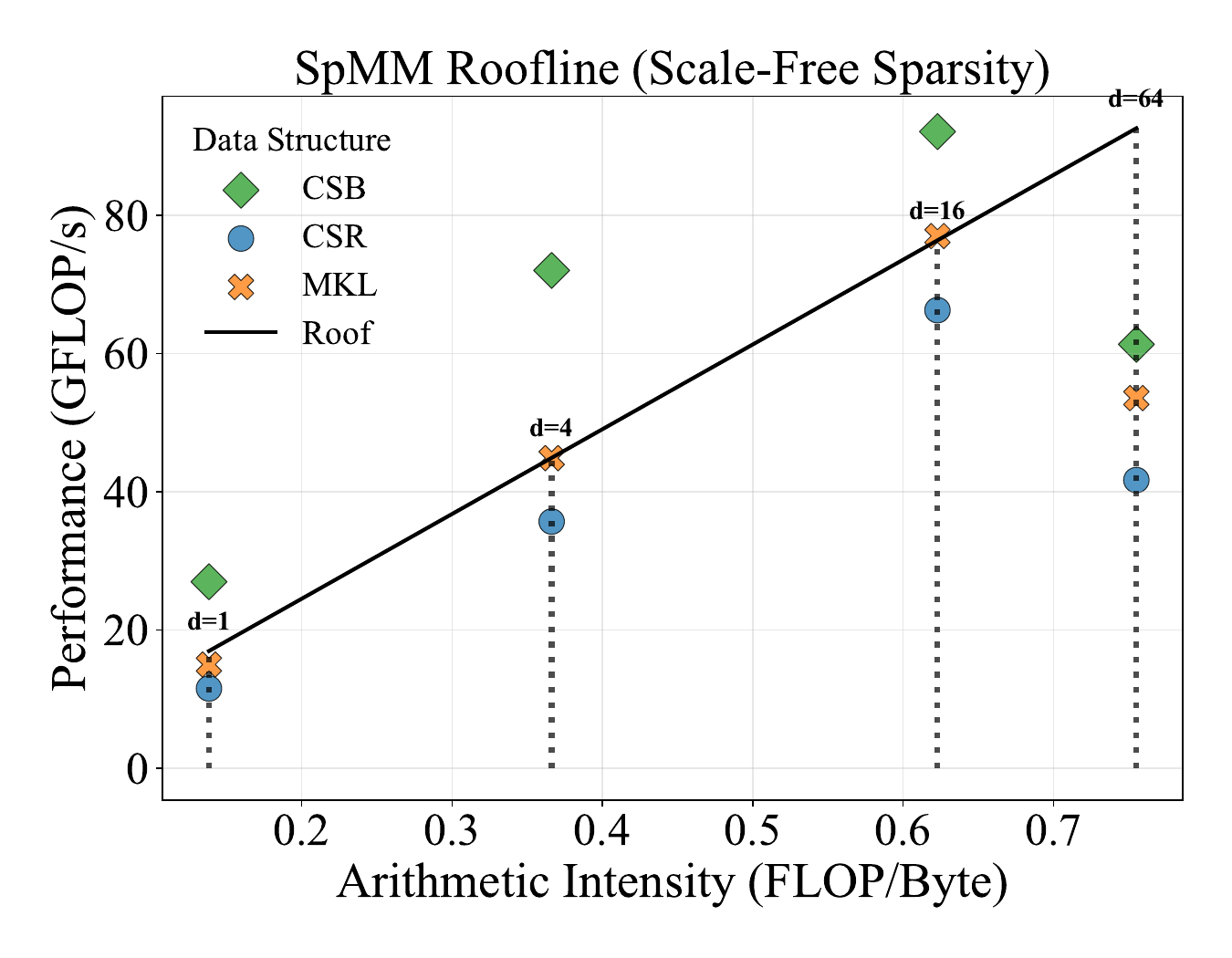}
\caption{Comparison of attained performance with our sparsity-aware Roofline model for four representative matrices, one from each sparsity pattern. Only the bandwidth-bound region of the Roofline is shown. In each plot, vertical lines indicate the theoretical arithmetic intensity for the corresponding sparsity pattern, as derived in Section~\ref{sec:roofline}. The intersection of each vertical line with the Roofline denotes the maximum attainable performance predicted by the model. Measured performance from CSR, CSB, and MKL implementations is overlaid using distinct symbols.}
\label{fig:roofline}
\end{figure*}

\subsection{Performance of SpMM implementations}
We first evaluate the performance of three SpMM implementations based on CSR and CSC data structures, as well as Intel MKL, for all matrices listed in Table~\ref{tab:matrices}. Experiments are conducted with dense matrix column dimensions $d = 1, 4, 16,$ and $64$, and the complete results are reported in Table~\ref{tab:spmm_k_multicol}. We then summarize the performance using four representative matrices, one from each sparsity pattern, and present these results in Figure~\ref{fig:spmm_all}.

We observe that uniformly random matrices consistently achieve the lowest performance, while scale-free matrices attain the highest performance across all three implementations and all values of $d$. These performance trends are validated in the next section using our sparsity-aware Roofline models. In addition, performance improves as the number of columns in $\mB$ increases; the lowest performance is observed at $d=1$, corresponding to SpMV, while the best performance is achieved near $d=32$ or $d=64$.






\subsection{Attained Performance Versus Roofline Model Predictions}
Figure~\ref{fig:roofline} illustrates the bandwidth-bound region of the Roofline model, with the bound determined by $P=\beta\cdot AI$.
The figure consists of four subplots, one for each sparsity pattern. Within each subplot, we report results for four dense matrix widths with $d = 1, 4, 16,$ and $64$. 
The closer the observed performance is to the bandwidth Roofline, the more accurately the model captures the performance behavior of SpMM. In the following, we compare the attained performance with our sparsity-aware Roofline models for each of the four sparsity patterns.

\subsubsection{Random Sparsity} 
For random sparsity, we observe that all three implementations achieve performance well below the Roofline, although CSB approaches the Roofline for $d=16$ and $d=64$. These results do not necessarily indicate suboptimal implementations; rather, they suggest the presence of limited data reuse even under random sparsity. In our random sparsity model, we assume no cache reuse, yielding a lower bound on arithmetic intensity. The empirical results, in which all implementations remain below the Roofline, are therefore consistent with this conservative assumption.

Nevertheless, random matrices may still offer some opportunities for cache reuse, which CSB is able to exploit more effectively, resulting in improved performance at larger $d$. In addition, random sparsity incurs high memory latency due to irregular accesses to rows of $\mB$. Because our Roofline model accounts only for bandwidth limitations and does not explicitly model memory latency, this effect may further explain the gap between the predicted maximum performance and the observed performance for random sparsity.


\subsubsection{Diagonal Sparsity}
Diagonal sparsity represents the opposite extreme, in which the model assumes substantial cache reuse and therefore provides an upper bound on arithmetic intensity. Because this model assumes near-perfect data reuse, an assumption that is rarely satisfied even for block-diagonal matrices, the observed performance of all implementations falls below the Roofline. Interestingly, the model is more accurate in the SpMV case, where modest deviations from an ideal diagonal structure do not significantly degrade performance.

While measured performance falls below the bound, this discrepancy is explained by deviations from an ideal diagonal structure in evaluated matrices. 
This leads to increased runtimes relative to the idealized model. Consequently, the banded sparsity roofline should be interpreted as a theoretical upper limit rather than a tight performance prediction for all diagonal-like matrices.

\subsubsection{Block Sparsity}
For matrices with block-structured sparsity, the performance attained by CSB closely approaches the memory-bandwidth Roofline, as predicted by the corresponding sparsity-aware model. 
This agreement validates the blocked sparsity model and demonstrates that CSB is well optimized for this structure, effectively exploiting spatial locality and data reuse.

\subsubsection{Scale-free Sparsity} 
For matrices with scale-free sparsity, the performance of CSR and MKL approaches the memory-bandwidth Roofline for small $d$, consistent with the near-optimal cache reuse predicted for hub nodes by the model.
However, for small $d$, CSB outperforms the predicted roof. This behavior suggests that the CSB data structure implicitly exploits localized block structure within the scale-free matrices that is not explicitly captured by the model, leading to additional reuse. Interestingly, for $d = 64$, all three data structures exhibit worse performance than $d=16$, and fall below the memory roof. 
This decline in performance could be due to the larger $d$ expanding the per-access working set for the vectors in $\mB$ and $\mC$. 
This means that even frequently reused rows associated with hot nodes no longer fit within the L1 cache. As a result, cache reuse deteriorates, effective memory traffic increases, and performance becomes limited by cache and memory hierarchy effects rather than arithmetic intensity alone.

The apparent violation of the bandwidth Roofline by CSB does not indicate a true exceedance of hardware limits, but rather reflects a mismatch between the assumed bandwidth ceiling and CSB’s effective access pattern. 
The Roofline plots use DRAM bandwidth measured by the STREAM benchmark, which characterizes long, streaming memory accesses. In contrast, CSB implicitly induces block-level locality that allows a substantial fraction of memory accesses to be serviced from cache. As a result, CSB operates under a higher effective bandwidth than the DRAM-only ceiling assumed by the model. This effect is most pronounced for small $d$, where rows of $\mB$ fit in cache and can be reused across block traversals. For larger $d$, the working set grows beyond cache capacity, cache reuse deteriorates, and CSB performance again falls below the bandwidth Roofline.

\section{Conclusion}
This work examined SpMM and performance evaluation through roofline modeling, with a focus on how matrix sparsity structure influences arithmetic intensity, memory traffic, and attainable performance. Rather than relying on a single model, we developed sparsity-aware roofline models and formulations that explicitly distinguish between block-structured, banded, scale-free, and uniformly random sparsity patterns. Our empirical results show that these structural differences fundamentally alter effective arithmetic intensity and performance bounds, and that a unified roofline model is insufficient to accurately explain observed SpMM behavior across diverse workloads. Using CSB, CSB, and MKL implementations, we showed that different sparsity structures and implementations lead to different scaling trends as $d$ increases.  Overall, our results suggest that accurate performance prediction and optimization of SpMM on modern architectures require sparsity-aware modeling rather than uniform assumptions about sparse matrices.

Our modeling approach also has several limitations. We made simplified assumptions about the sparsity structure, particularly for scale-free and blocked matrices. As a result, it is difficult to determine whether the gap between observed performance and peak performance arises from modeling inaccuracies or implementation inefficiencies.
In addition, our model does not adequately capture cache behavior and ignores memory latency effects. We acknowledge that both factors should be incorporated into a more realistic model to better reflect the true performance of SpMM.
Finally, the Roofline model itself may not be fully sufficient for capturing SpMM performance due to the irregular nature of sparse computations. Therefore, alternative modeling approaches may be necessary to more completely understand and characterize the performance of SpMM.

\section*{Acknowledgments}
This research was funded in part by DOE grants DE-SC0022098 and DE-SC0023349 and by NSF grants PPoSS CCF 2316233 and OAC-2339607.



\captionsetup[table]{
  width=\linewidth,
  justification=centering
}

\bibliographystyle{IEEEtran}
\bibliography{Ref}

\appendix

\section{Scale-free graphs}

{\bf Derivation of the hub edge fraction for scale-free graphs}
Assume that the degree distribution $p(k)$ of a graph follows a power law, and $k_{\min}$ is the minimum degree for which the power law holds. Then the degree distribution is expressed as:

\begin{equation}
p(k) = C k^{-\alpha}, \quad k \ge k_{\min},
\end{equation}
where $C$ is a normalization constant. 
The power-law exponent is typically $2 < \alpha < 3$ for real-world networks.

The total number of nonzeros (equivalently, total degree mass) is proportional to the sum of degrees,
\begin{equation}
nnz \;\propto\; \sum_k k \;\approx\; \int_{k_{\min}}^{\infty} k\, p(k)\, dk
= C \int_{k_{\min}}^{\infty} k^{1-\alpha} \, dk.
\end{equation}
For $\alpha > 2$, this integral converges and evaluates to
\begin{equation}
\int_{k_{\min}}^{\infty} k^{1-\alpha} \, dk
= \frac{k_{\min}^{2-\alpha}}{\alpha-2}.
\end{equation}

Similarly, the contribution of hub nodes with degree at least $K$ to the total number of nonzeros is
\begin{equation}
nnz_{\mathrm{hub}} \;\propto\;
\int_{K}^{\infty} k\, p(k)\, dk
= C \int_{K}^{\infty} k^{1-\alpha} \, dk
= \frac{C}{\alpha-2} K^{2-\alpha}.
\end{equation}

Taking the ratio of these two quantities yields
\begin{equation}
\frac{nnz_{\mathrm{hub}}}{nnz}
=
\frac{K^{2-\alpha}}{k_{\min}^{2-\alpha}}
=
\left( \frac{K}{k_{\min}} \right)^{2-\alpha}.
\end{equation}

This expression shows that, for scale-free graphs with $\alpha$ close to $2$, a small number of high-degree hub nodes can account for a disproportionately large fraction of nonzeros. In the context of SpMM, this concentration directly translates to increased reuse of dense matrix rows associated with hub vertices and, consequently, higher effective arithmetic intensity than predicted by random sparsity models.

\paragraph{Hub edge fraction as a function of hub ratio}
If hubs are defined as the top fraction $f$ of nodes by degree, then from the complementary cumulative distribution of the power-law,
\begin{equation}
f \;\approx\; \Pr(k \ge K)
\;\approx\;
\left( \frac{K}{k_{\min}} \right)^{1-\alpha},
\end{equation}
which implies
\begin{equation}
K \approx k_{\min} \, f^{-1/(\alpha-1)}.
\end{equation}

Substituting this expression for $K$ into the ratio of nonzeros contributed by hub nodes,
\begin{equation}
\frac{nnz_{\mathrm{hub}}}{nnz}
=
\left( \frac{K}{k_{\min}} \right)^{2-\alpha},
\end{equation}
yields
\begin{align}
\frac{nnz_{\mathrm{hub}}}{nnz}
&=
\left( f^{-1/(\alpha-1)} \right)^{2-\alpha} \\
&=
f^{-(2-\alpha)/(\alpha-1)} \\
&=
f^{(\alpha-2)/(\alpha-1)}.
\end{align}

For $\alpha>2$, the exponent $(\alpha-2)/(\alpha-1)$ is positive, indicating that even a small fraction of hub nodes can account for a substantial fraction of nonzeros. In particular, when $\alpha$ is close to $2$, the edge mass is highly concentrated among hubs, which has important implications for data reuse and arithmetic intensity in SpMM.

For example, if $\alpha = 2.2$, then $\frac{\alpha-2}{\alpha-1} = 0.2/1.2 \approx 0.167$. Then, $f=1\%$ gives us $\frac{nnz_{\mathrm{hub}}}{nnz} \approx 0.46$.




\end{document}